\documentclass[nologo,11pt,a4paper]{ETHpaper}
\usepackage{amsmath, amssymb}
\usepackage{graphicx}
\usepackage{subfigure}
\usepackage{scalefnt}
\usepackage{enumitem}

\usepackage{tikz}
\usepackage{pgf}
\usepackage{pgfplots}
\usepackage{scalefnt}

\usetikzlibrary{plotmarks,arrows,shapes,positioning}

\newcounter{Temporal Networks: }

\begin{document}

\title{Betweenness Preference: Quantifying Correlations in the Topological Dynamics of Temporal Networks}
\author{Ren\'{e} Pfitzner, Ingo Scholtes, Antonios Garas, Claudio J. Tessone, Frank Schweitzer}
\address{
  Chair of Systems Design, ETH Zurich\\
  Weinbergstrasse 56/58, 8092 Zurich, Switzerland \\
  \texttt{\{rpfitzner,ischoltes,agaras,tessonec,fschweitzer\}@ethz.ch}
}

\maketitle
\begin{abstract}
We study correlations in temporal networks and introduce the notion of betweenness preference.
It allows to quantify to what extent paths, existing in time-aggregated representations of temporal networks, are actually realizable based on the sequence of interactions.
We show that betweenness preference is present in empirical temporal network data and that it influences the length of shortest time-respecting paths.
Using four different data sets, we further argue that neglecting betweenness preference leads to wrong conclusions about dynamical processes on temporal networks.
\end{abstract}
Recent works have argued that properties of dynamical processes evolving on complex networks change profoundly when the dynamics of the network topology is taken into account.
For a number of empirical temporal networks obtained from time-stamped contact data, simulations have shown that their topological dynamics can both slow down~\cite{IseSteBarCatPinBro2011,Blonder2011,KarKivPanKasKerBarSar2011} or speed up spreading processes~\cite{Rocha2011}.
At the same time, it has been observed that, compared to time-aggregated topologies, the exploration dynamics of random walks in temporal networks is significantly slower~\cite{StaBarBarPas2012}.
Furthermore, it has been shown that network dynamics alone can give rise to collective phenomena like synchronization~\cite{Tessone2012}.
These observations have generated significant interest in the mechanism underlying these phenomena.
A series of recent works focused on the influence of inter-event time distributions and temporal correlations in the time series of interactions~\cite{IriMor2009,KarKivPanKasKerBarSar2011,StaBarBarPas2012,PerGonPasVes2012,Miritello2011}.
Bursty activity patterns of nodes have been identified as one possible source that slows down spreading~\cite{KarKivPanKasKerBarSar2011} and random walk processes~\cite{PerBarMocGonPasVes2012}.
Similarly, bursty node activities have been suggested to slow down information diffusion, particularly when the diffusion process is initiated in phases of low activity~\cite{Panisson2012}.
Furthermore, for a number of social contact networks, it has been shown that heterogeneous inter-event times increase the length of time-respecting paths~\cite{PanSar2011}.
Apart from inter-event time distributions, it has been argued that link appearance frequencies and their correlation with community structures are another characteristic of temporal networks that can slow down spreading dynamics~\cite{KarKivPanKasKerBarSar2011}.
Another line of research is concerned with the study of \emph{temporal motifs}~\cite{KovKarKasKersar2011,QZhao2010}, i.e.~whether there are classes of frequently occurring temporal contact patterns.
It was shown that the presence of certain temporal motifs (like e.g.~``chains'' of consecutive edges continuing time-respecting paths) can \emph{decrease} the length of time-respecting paths, thus speeding up spreading processes~\cite{PanSar2011}.

Going beyond previous works, in this Letter we study \emph{betweenness preference}, which captures the over- or under-representation of particular time-respecting paths passing through nodes.
This \emph{temporal-topological} feature is neither visible in the weighted time-aggregated network, nor can it be attributed to inter-event time distributions, bursty node dynamics or the statistics of temporal motifs.
Our study is motivated by the idea that in many real-world networks nodes contact other nodes based on previous contacts~\cite{Miritello2011}.
As a simple example consider the influence of context in information dissemination: e-mails received by work-related contacts are more likely to be forwarded to a work-related subset of social contacts.
Here we study the influence of such special classes of dynamical contact patterns.
We argue that betweenness preference, i.e.~the tendency of nodes to preferentially connect - in a temporal sense - particular pairs of neighbors, (i) is not captured in the time-aggregated network, (ii) is present in empirical temporal network data, (iii) changes the topology of time-respecting paths, and (iv) critically influences dynamical processes evolving on temporal networks.

\begin{figure}[t]
\includegraphics[width=\columnwidth]{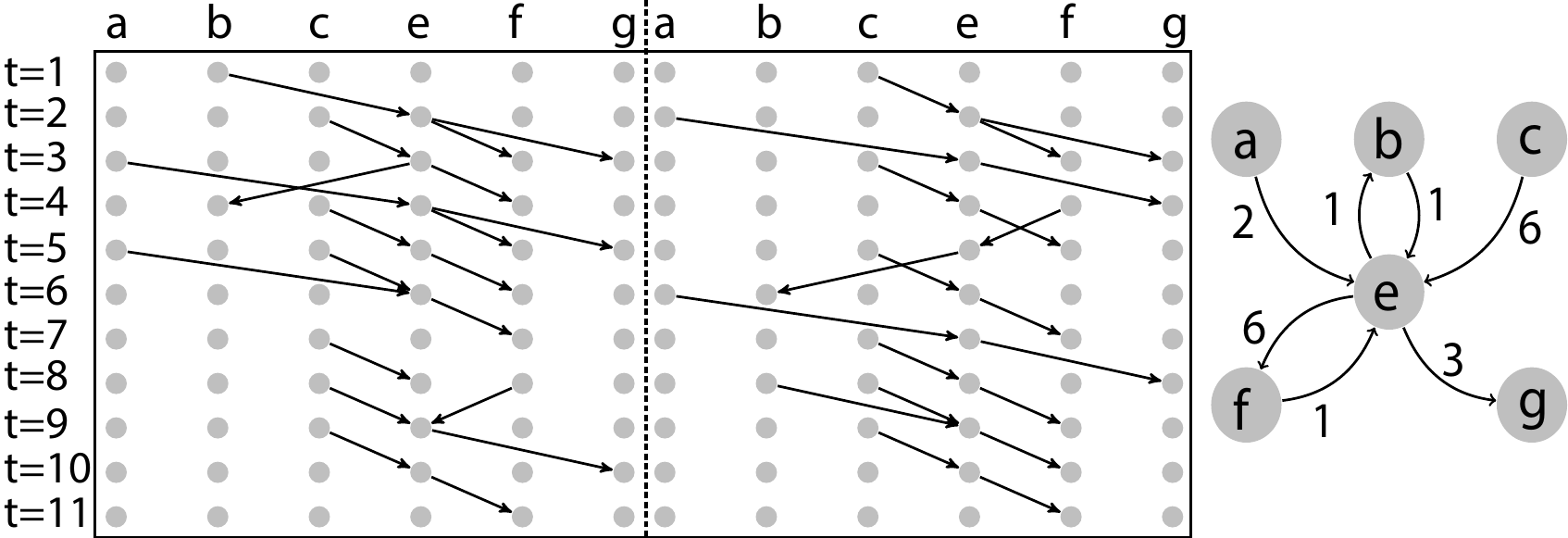}
\caption{Time-aggregated weighted network $G_{Agg}$ (right) and time-unfolded network of two different temporal networks $G_{Dyn_1}$ and $G_{Dyn_2}$ (left), both of which are consistent with $G_{Agg}$.}
\label{fig:example}
\end{figure}

A temporal network is defined as a tuple consisting of a set of nodes $v \in \mathcal{V}$ as well as a set of \emph{events}: $e(v, w, t, l \, \Delta t)\in \mathcal{E}$.
An event is an interaction between two nodes $v$ and $w$, starting at time $t$ and with a duration $l \, \Delta t$  relative to some smallest unit of discrete time $\Delta t$ (for simplicity, we assume $\Delta t=1$).
Based on time-stamped edges and a discrete notion of time, we construct a \emph{flow-preserving} static representation of temporal networks by \textit{unfolding time} into an additional topological dimension.
This construction serves as the basis for our models, and we call it a \textit{time-unfolded network}.
Time-unfolded networks of two different temporal networks are illustrated in Fig.~\ref{fig:example}.
In the resulting temporal unfolding, we indicate the presence of a possible flow event by an edge $(v_t, w_{t+1})$, while replacing the original node set $\mathcal{V}$ by a set $\mathcal{V}'$ of \emph{temporal copies} of nodes $v_t$ where $v \in \mathcal{V}$ and $t \in \{ 0, 1, \ldots, L  \}$ for an observation period of duration $L$,
similar to~\cite{ParDicCohStaHav2010,KimAnd2012,MitTabRot2012}.
As shown in Fig.~\ref{fig:example} (left), the two different temporal networks are the same in the time-aggregated representation $G_{Agg}$ (right).
In $G_{Agg}$ edge weights indicate the number $l$ of time steps in which an edge was active throughout the observation period.
In analogy to Statistical Mechanics, one might think of such a time-aggregated network as a macro-state which is compatible with different temporal networks, i.e.~micro-states.
\paragraph*{Betweenness Preference.}
An important aspect when studying dynamical processes like diffusion or synchronization on \textit{static, time-aggregated} networks is that one assumes \emph{transitive paths}.
However, this transitivity does not necessarily hold in a temporal network that gives rise to the respective time-aggregated network~\cite{HolSar2012}.
To illustrate this fact, consider the time-aggregated network $G_{Agg}$ depicted in Fig.~\ref{fig:example}.
In a static system, information could spread in a transitive way from $b$ via $e$ to $g$.
In a temporal network the order in which edges appear imposes an additional constraint: information can only flow along \emph{time-respecting paths}~\cite{HolSar2012}.
Hence, in a temporal network underlying $G_{Agg}$, information can only flow from $b$ to $g$ if the connection $(b,e)$ is \textit{followed} by a connection $(e,g)$.
Thus, even though the links $(b,e)$ and $(e,g)$ are present in both temporal networks (Fig.~\ref{fig:example} left), a time-respecting path between $b$ and $g$ only exists in the left example.
To quantify this transitivity-limiting property of temporal networks, we study whether certain time-respecting paths are preferentially realized as compared to the time-aggregated perspective.
Focusing on elementary building blocks of time-respecting paths, we particularly study \emph{two-paths}, i.e.~a path of length two, representing two consecutive edge activations interconnecting three nodes.
The statistics of the ensemble of realized two-paths will reveal to what extent path-transitivity holds in the temporal network.
Based on the time-unfolded representation of temporal networks, we define the elements of a per-node \emph{betweenness preference matrix} $\mathbf{B}^v(t)$ as follows:
\begin{equation}
    B^v_{sd}(t) := \begin{cases}
        1, & \text{if } (s_{t-1}, v_t) \in \mathcal{E}  \text{ and } (v_t, d_{t+1}) \in \mathcal{E}\\
        0, & \text{otherwise } 
    \end{cases}.
\label{eqn:BPM}
\end{equation}
Each matrix element $B^v_{sd}(t)$ captures whether node $v_t$ in a time-unfolded network was in between a source $s_{t-1}$ and a destination $d_{t+1}$ on a two-path $(s_{t-1}, v_t)\rightarrow (v_t, d_{t+1})$.
This definition builds on a notion of time-respecting paths comprised of edge activations following each other \textit{immediately}, can be relaxed though by including some notion of memory.
Based on definition \eqref{eqn:BPM}, we define the elements of a \textit{time-aggregated betweenness preference matrix} $\mathbf{B}^v$:
\begin{equation}
    B^v_{sd} := \sum_t{B^v_{sd}(t) \cdot \left[ \sum_{s'd'} B^v_{s'd'}(t) \right]^{-1}},
    \label{eqn:B_Matrix}
\end{equation}
and a \textit{normalized betweenness preference matrix} $\mathbf{P}^v$:
\begin{equation}
    P^v_{sd} := B^v_{sd} \cdot \left[ \sum_{s'd'} B^v_{s'd'} \right]^{-1}.
    \label{eqn:P_Matrix}
\end{equation}
Essentially, $P^v_{sd}$ is the probability distribution of the two-paths $(s_{t-1},v_{t})\rightarrow (v_{t},d_{t+1})$ over all $t$.
We use this to quantify to what extent $v$ exhibits a preference to interconnect particular pairs of source and target nodes.
Based on the concept of mutual information, we define a \textit{betweenness preference measure} as
\begin{equation}
    I^v(S;D) := \sum_{\substack{d \in D\\s \in S}} P^v_{sd} \log_{2} \left( \frac{P^v_{sd}}{P^v(s) P^v(d)} \right),
    \label{eqn:I_Measure}
\end{equation}
where $P^v(s) = \sum_d{P^v_{sd}}$ and $P^v(d) = \sum_s{P^v_{sd}}$.
In general, $I^v(S;D)$ captures to what extent the knowledge of the source $s$ of a time-respecting path through $v$ determines the next step $d$.
Or, within a context of information flow, it measures how selective $v$ is in mediating information preferentially between certain pairs of nodes $s$ and $d$.
We note that this measure is minimal if the random variables $S$ and $D$ are independent.
This allows us to calculate the matrix elements $P^v_{sd}$ resulting in $I^v(S;D)=0$ solely based on the underlying static, time-aggregated network with edge weights $w_{ij}$:
\begin{equation}
  \hat{P}^v_{sd} := p^v(s) \cdot p^v(d),
  \label{eq:uncorrelated}
\end{equation}
where $p^v(s) = w_{sv} \left[ \sum_{s'}{w_{s'v} } \right]^{-1}$ and $p^v(d) = w_{vd} \left[ \sum_{d'}{w_{vd'} } \right]^{-1}$.

We now introduce a configuration model to generate temporal networks that are members of an ensemble defined by a given set of betweenness preference matrices.
We limit ourselves to a subset of all possible realizations, in which exactly one edge is active per time step and which \emph{only} consist of two-paths.
The model creates a temporal network with given betweenness preference matrices as follows:
First, we define the number of two-paths $N_2$ to be realized.
Second, we draw a random two-path $(s,v)\rightarrow (v,d)$ according to $p(s,v,d):=B^v_{sd} / \sum_{v,s,d} B^v_{sd}$.
Third, we create temporal network edges $(s_t,v_{t+1})$ and $(v_{t+1},d_{t+2})$.
We increment $t=t+3$ and $n_2=n_2+1$ (number of realized two-paths) and repeat this procedure until $n_2=N_2$.
Having \textit{empirical} temporal network data available, we use this model to create micro-states that (i) preserve the betweenness preference distribution (in an infinitely long sequence), (ii) have the same macro-state, and (iii) destroy all other correlations (such as bursty activity patterns).
We call the so constructed temporal network \textit{betweenness preference preserving case}.
\begin{figure}[t]
\includegraphics[width=0.9\columnwidth]{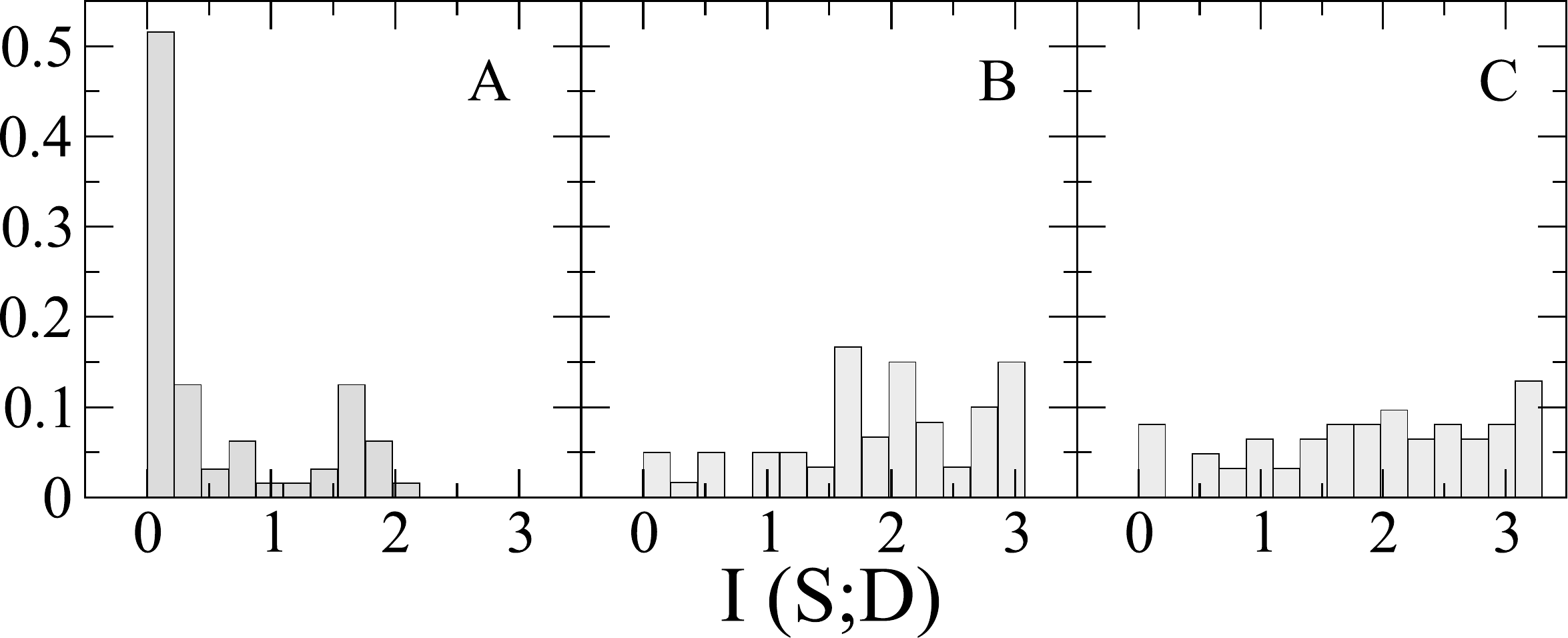}
\caption{Betweenness preference distributions for nodes of (A) 200 temporal networks of the uncorrelated case, (B) the empirical RealityMining data and (C) 200 temporal networks of the $\mathbf{P}^v$ preserving case. All temporal networks have the same duration of $N_2=5\cdot10^4$ two-paths. (A) and (C) are based on the same empirical data as (B).}
\label{fig:new_combined}
\end{figure}
Using this configuration model, we also construct micro-states with low betweenness preference based on a given macro-state, by using the probability $\hat{P}^v_{sd}$ defined in eq.~\eqref{eq:uncorrelated}.
We call a temporal network created in such a way \textit{uncorrelated case}, since it only preserves the macro-state (the weighted aggregate network), but destroys betweenness preference.
Precisely, the uncorrelated case has betweenness preference that is expected from a random micro-state of \textit{finite duration} (in a temporal network of infinite duration, the model reproduces the limiting case of $I^v(S;D)=0$).
\paragraph*{Empirical Results.}
To demonstrate that betweenness preference is an important property in real-world data sets, we use a one-week subset of the empirical contact sequence (Sept.~8th to 15th, 2004) from the RealityMining Project (RM)~\cite{RealityMining}, featuring 64 individuals with $20,000$ recorded interactions and a granularity of $5$ minutes.
Fig.~\ref{fig:new_combined} shows the distribution of betweenness preference in (A) the uncorrelated case, (B) the original data sample and (C) the betweenness preference preserving case.
The uncorrelated (A) and betweenness preference preserving (C) cases were created using the betweenness preference matrices of the RM data, utilizing the configuration model described above.
As expected, the uncorrelated case (A) shows a spike around $I(S;T)=0$, indicating small betweenness preference for most nodes.
The theoretical expectation of $I(S;T)=0$ for all nodes in the uncorrelated case is not realized due to finite duration of the temporal sequence.
Analyzing the betweenness preference distribution of the empirical temporal network (B), one realizes that it is very different from the one in (A).
The distribution in (B) is rather broad, with an average $\langle I(S;T) \rangle=1.9$ and a median $Q_{0.5}(I(S;T))=1.99$, as compared to $\langle I(S;T) \rangle=0.58$ and $Q_{0.5}(I(S;T))=0.21$ in (A).
Performing a two-sided Kolmogorov-Smirnov (KS) test, we can reject the hypothesis that distribution (B) is identical to distribution (A) ($p<10^{-9}$).
This shows that there is a pronounced amount of betweenness preference in the empirical contact sequence, whereas it is mostly absent in the uncorrelated case.
Panel (C) shows the betweenness preference distribution of the betweenness preference preserving case.
With $\langle I(S;T) \rangle=1.91$ and median $Q_{0.5}(I(S;T))=2.02$, distributions (B) and (C) are very similar.
Since we create the model in a statistical fashion based on the normalized betweenness preference matrix $\mathbf{P}^v$, the two distributions are not completely identical due to finite $N_2$.
Performing the two-sided KS test, we cannot reject the hypothesis that the two distributions are identical ($p=0.68$).
Hence, in this case the model preserves the betweenness preference of the real network, whereas all other correlations (e.g.~bursty node activities) are destroyed by construction.

\begin{figure}[t]
\subfigure{
    \includegraphics[width=0.33\columnwidth]{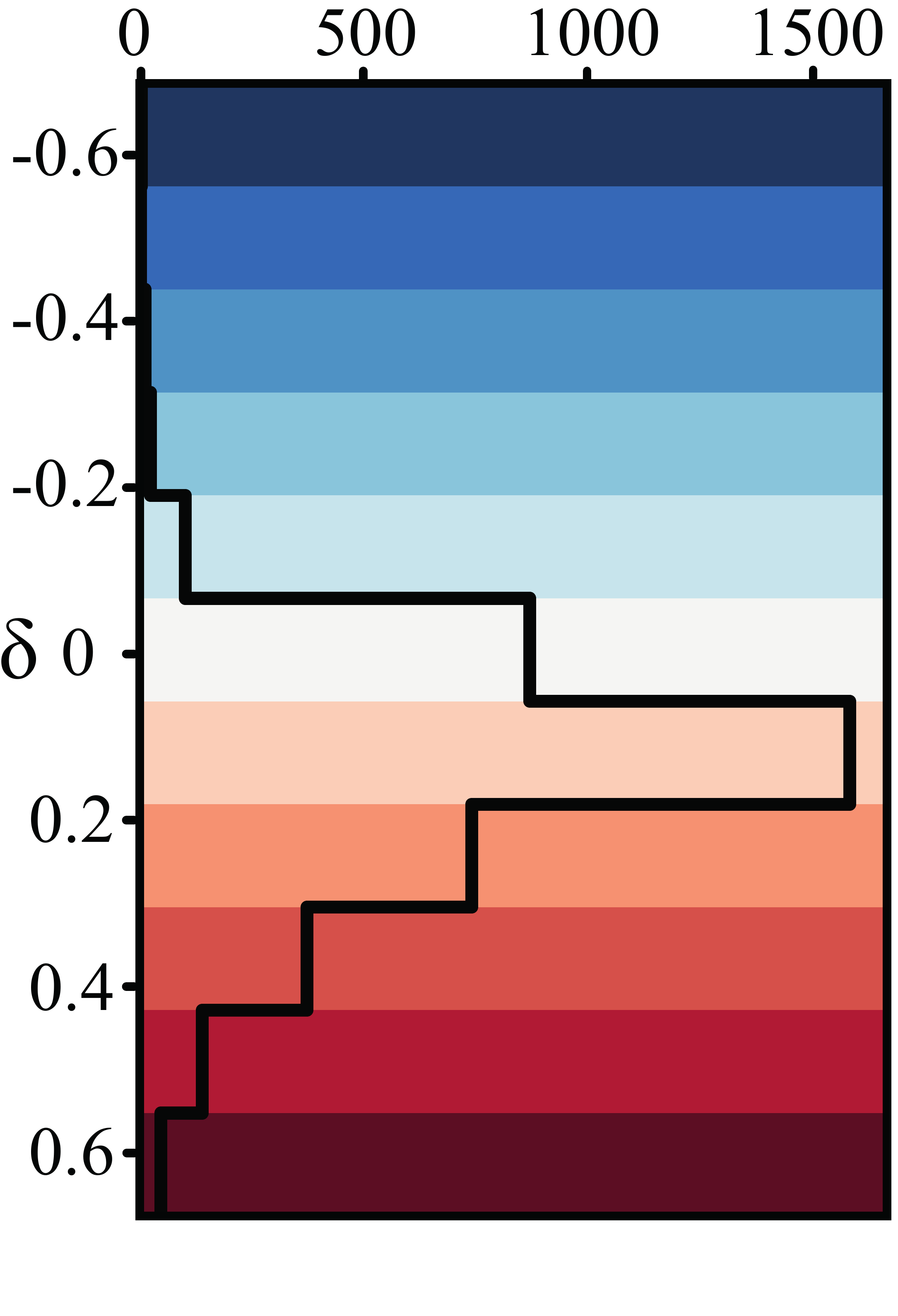}
    }
\subfigure{
    \includegraphics[width=0.49\columnwidth]{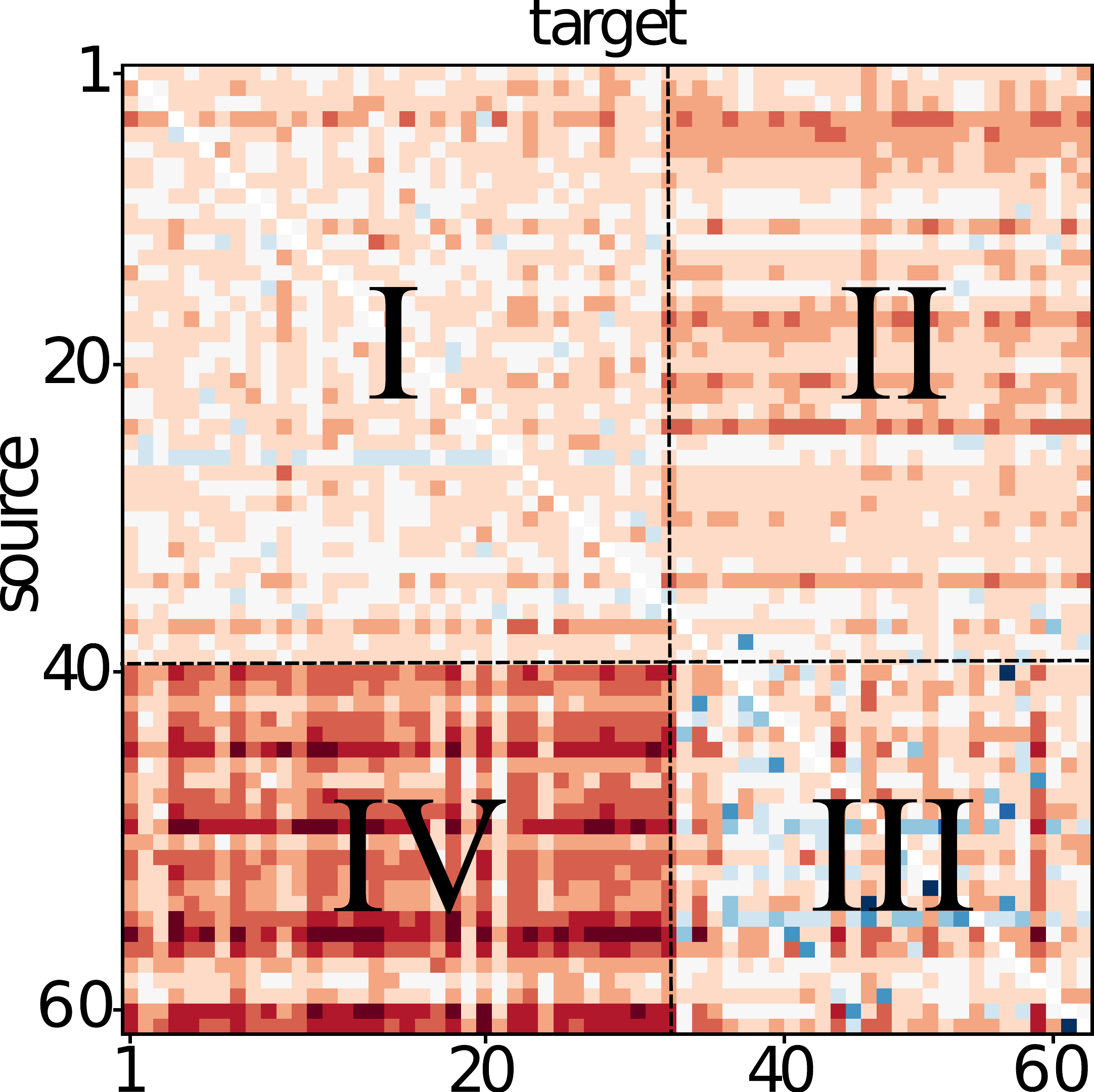}
    }
\caption{(Color online) Relative difference $\delta$ between the lengths of fastest time-respecting paths in betweenness preference preserving and uncorrelated temporal networks constructed from the RealityMining data set.
Each cell of the matrix represents the length difference for one pair of nodes, averaged over $200$ realizations of betweenness preference preserving and uncorrelated temporal networks.
Rows and columns are sorted in the same way.
Notation I-IV is included to highlight the block structure of the matrix.
A histogram of relative differences is overlayed in the color legend (
left panel).
}
\label{fig:pathlengths}
\end{figure}

An important aspect of betweenness preference is that it influences \emph{transitivity} and, thus, changes the \emph{topology of time-respecting paths} in temporal networks.
Of particular interest in the study of temporal networks is the notion of \emph{fastest time-respecting paths}.
A fastest time-respecting path (ftrp) between nodes $s$ and $v$ is defined as the path by which information from node $s$ \emph{first} reaches node $v$~\cite{HolSar2012}.
To quantify the transitivity-limiting effect of betweenness preference, i.e.~the under- or over-representation of \emph{particular two-paths}, we study ftrp's in temporal networks, where two-paths are the only means of information flow, as follows.
We measured the length of the ftrp between all pairs of nodes for uncorrelated ($L_u(s,v)$) and betweenness preference preserving realizations ($L_p(s,v)$), generated by our configuration model.
For $s=v$, we set zero path length.
We compute the relative length difference $\delta(s,v)=\left(L_p(s,v)-L_u(s,v)\right)/L_u(s,v)$ and average over several realizations.
%
Results for the RM data set are shown in Fig.~\ref{fig:pathlengths}.
The left panel shows the distribution of $\delta(s,v)$ which exhibits a clear shift to positive values.
Moreover, the null hypothesis that the distribution of length of fastest time-respecting paths in the betweenness preference preserving and uncorrelated case are the same can be rejected (two-sided KS test yields $p<10^{-15}$).
Thus we conclude that betweenness preference has a profound effect on the topology of fastest time-respecting paths in temporal networks ($\langle \delta(s,v) \rangle = 0.08$, $Q_{0.5}(\delta(s,v)) = 0.06$).
In Fig.~\ref{fig:pathlengths}, right panel, we present the matrix $\delta(s,v)$, exhibiting a pronounced block structure.
In particular, high values of $\delta(s,v)$ occur in the off-diagonal regions IV and II (albeit less pronounced in the latter).
This result implies the existence of a temporal community structure induced by the existence of betweenness preference in the RM data set: fastest time-respecting paths are up to $60\%$ \emph{longer} between communities (blocks II and IV), whereas the prolonging effect almost vanishes within the communities (blocks I and III).
For some pairs of nodes even the opposite effect is apparent in region III: betweenness preference selectively \textit{shortens} fastest time-respecting paths. 
\paragraph*{Spreading Dynamics in Temporal Networks.}
%
Given this result on the effect of betweenness preference on time-respecting paths, we now study the paradigmatic SI (Susceptible-Infected) epidemic model to quantify the effect of betweenness preference on dynamic processes evolving on temporal networks.
To exclusively quantify the impact of betweenness preference correlations, we compare spreading dynamics on temporal networks created with the uncorrelated case (A), and the betweenness preference preserving case (C) of our configuration model.
More specifically, we study four different data sets:
\textbf{(AN)} social interactions in ant colonies
~\footnote{We consider the temporal network for one ant colony, consisting of $81$ individuals and $475$ interactions \cite{Blonder2011}.}; 
\textbf{(CN)} a large temporal network of scientific co-authorships~\footnote{Collected from a scholarly database, it consists of $10^4$ authors and $1.8\cdot 10^6$ co-authorship relations. It spans the complete history of papers published within the domain ``Condensed Matter Physics'' from 1926 to 1999.}; 
\textbf{(RM)} the RealityMining data set introduced earlier;
and \textbf{(SN)} a medium-sized synthetic network~\footnote{The network has similar modularity, cluster structure, and density as the RM data, but with $1000$ nodes, and the nodes additionally have artificially high betweenness preference (a procedure based on \cite{NewGir2004}).}.
\begin{figure}[t]
\includegraphics[width=0.9\columnwidth]{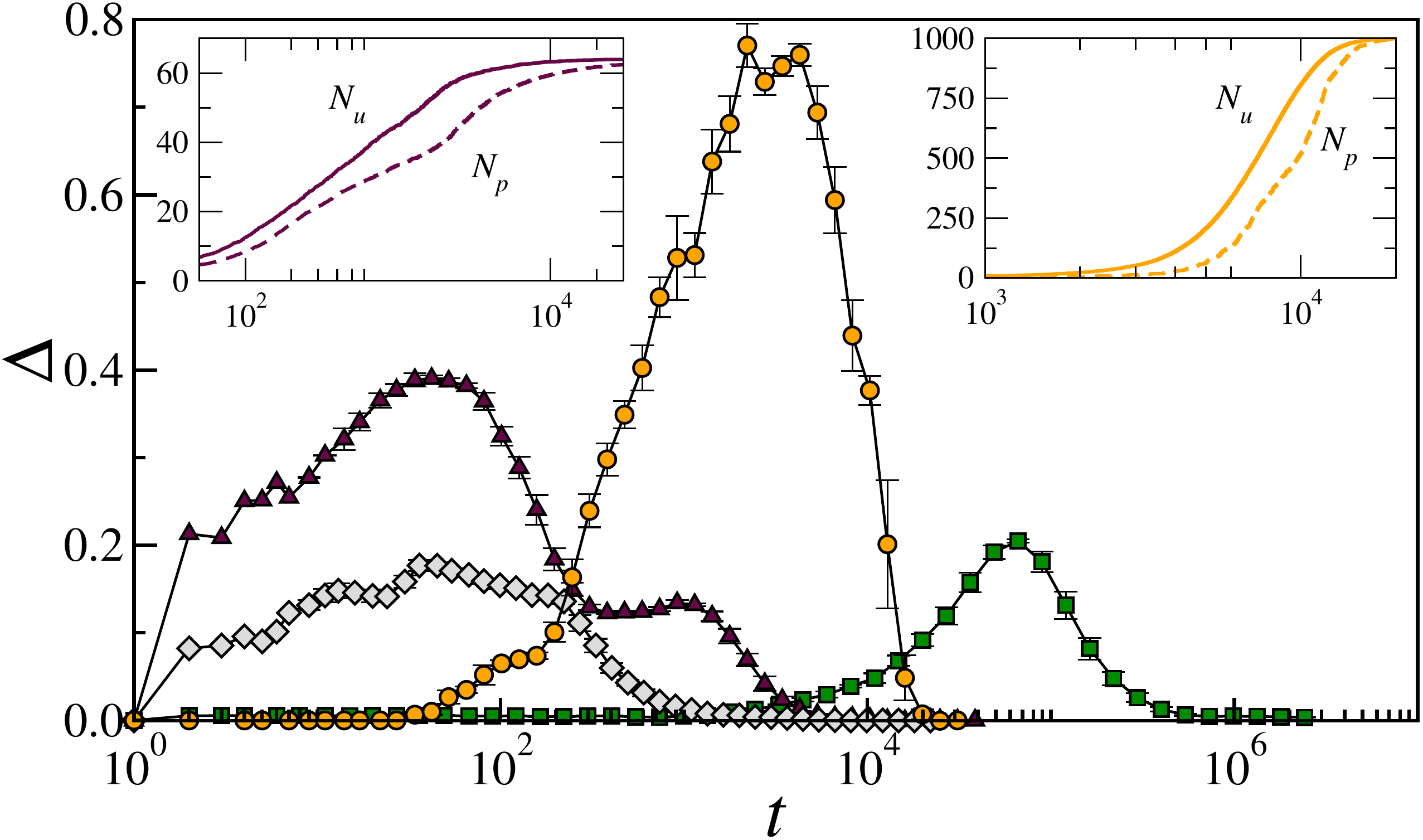}
\caption{(Color online) SI-spreading dynamics for temporal networks 
of RealityMining data (RM, purple $\triangle$), the ant network (AN, grey $\diamondsuit$), the co-authorship network (CN, green $\square$) and the $1000$ node synthetic network (SN, orange $\bigcirc$). Main figure shows time evolution of average relative differences $\Delta$ ($10^4$ realizations for CN, 200 for the others) in SI spreading dynamics between the uncorrelated and the betweenness preference preserving cases. Insets (left: RM, right: SN) show time evolution of number of infected nodes for the uncorrelated ($N_u$) and betweenness preference preserving case ($N_p$). Error bars represent standard deviations (when not shown, error bars are smaller than symbol size).}
\label{fig:si_spreading}
\end{figure}
To start the contagion process, in all simulations the first appearing node has been infected and the infection probability set to $p=1$.
The insets of Fig.~\ref{fig:si_spreading} show the evolution of the number of infected individuals for the uncorrelated and the betweenness preference preserving cases for the RM and SN data sets.
The number of infected individuals clearly follows a typical S-shaped curve (however, notice the semi-logarithmic scaling) in the uncorrelated $N_{u}$ as well as the betweenness preference preserving $N_{p}$ cases.
The slopes in the middle part of the infection dynamics are however clearly different, indicating slower spreading in the temporal network with non-vanishing betweenness preference $N_{p}$.
This slow-down is indicated by the time-to-saturation which, in the RM data (Fig.~\ref{fig:si_spreading}, left inset), is one order of magnitude larger for $N_{p}$.
To substantiate this important effect of large betweenness preference, the main panel shows the time evolution of the relative difference of infected individuals in the uncorrelated and the betweenness preference preserving cases, i.e.~$\Delta=(N_{u}(t)-N_{p}(t)) / N_{u}(t)$.
Our results clearly show that the uncorrelated model significantly overestimates the average number of infected individuals - for RM 
at times up to $\approx 40\%$, for AN 
and CN 
up to $\approx 20\%$.
Results for the synthetic network with artificially high betweenness preference (Fig.~\ref{fig:si_spreading}, circles and right inset) confirm these findings and show evidence that in large systems high betweenness preference can have an even more pronounced effect on spreading processes: the slow-down of spreading processes can be as large as $\approx 80\%$.
Interpreting $\Delta$ as the \textit{error} made when not accounting for betweenness preference correlations, it becomes obvious that taking a time-aggregated perspective on temporal networks, and hence neglecting betweenness preference, can lead to significantly misleading statements about the properties of dynamical processes evolving on networks with dynamic topology.
Betweenness preference, captured in terms of the measure introduced in eq.~\eqref{eqn:I_Measure}, quantifies this potential pitfall and helps to decide whether time-respecting paths are statistically distributed as expected from a weighted time-aggregated perspective.
Additionally, \emph{betweenness preference matrices} (eqs.~\eqref{eqn:B_Matrix} and \eqref{eqn:P_Matrix}) allow (i) to study the over- or under-representation of \emph{particular} time-respecting paths passing through nodes and (ii) to define proxy models that reproduce the temporal-topological dynamics of empirical temporal networks.
\paragraph*{Acknowledgements.}
I.S. acknowledges support by the SNF through grant CR12I1\_125298. C.J.T. and A.G. acknowledge support by the SNF through grant 100014\_126865. A.G. and F.S. acknowledge support by the EU-FET project MULTIPLEX 317532.
\bibliographystyle{plain}
\bibliography{paper}
\end{document}